\documentclass[preprint,superscriptaddress,aps]{revtex4}
\usepackage{amsfonts,amsmath,amssymb}
\usepackage{amsfonts}
\usepackage{graphics}
\usepackage{graphicx}
\usepackage{epsf}

\newcommand{\bea}{\begin{eqnarray}}
\newcommand{\eea}{\end{eqnarray}}

\begin{document}

\title{Statistical properties of linear Majorana fermions}

\author{F. C. E. Lima\footnote{E-mail: cleiton.estevao@fisica.ufc.br}}
\affiliation{Universidade Federal do Cear\'{a} (UFC), Departamento do F\'{i}sica - Campus do Pici, Fortaleza, CE, C. P. 6030, 60455-760, Brazil.}

\author{A. R. P. Moreira\footnote{E-mail: allan.moreira@fisica.ufc.br}}
\affiliation{Universidade Federal do Cear\'{a} (UFC), Departamento do F\'{i}sica - Campus do Pici, Fortaleza, CE, C. P. 6030, 60455-760, Brazil.}

\author{L. E. S. Machado\footnote{E-mail: lauramachado@fisica.ufc.b}}
\affiliation{Universidade Federal do Cear\'{a} (UFC), Departamento do F\'{i}sica - Campus do Pici, Fortaleza, CE, C. P. 6030, 60455-760, Brazil.}

\author{C. A. S. Almeida\footnote{E-mail: carlos@fisica.ufc.br}}
\affiliation{Universidade Federal do Cear\'{a} (UFC), Departamento do F\'{i}sica - Campus do Pici, Fortaleza, CE, C. P. 6030, 60455-760, Brazil.}

\begin{abstract}
A Majorana fermion is the single fermionic particle that is its own antiparticle. Its dynamics is determined by the Majorana equation, where the spinor field is by definition equal to its charge-conjugate field. In this paper, we investigated Shannon's entropy of linear Majorana fermions to understand how this quantity is modified due to an external potential of the linear type linear. Subsequently, we turn our attention to the construction of an ensemble of these Majorana particles to study the thermodynamic properties of the model. Finally, we show how Shannon's entropy and thermodynamic properties are modified under the linear potential action.
\end{abstract}%

\maketitle

\section{Introduction}\label{sec1}

In 1937, Ettore Majorana studying a relativistic version of quantum models, proposed the equation known as the Majorana equation \cite{Majorana}. Through his theory, it is known that an electrically neutral particle, also known as Majorana fermions, is similar to its antiparticle. Due to this property, the Majorana spinor has half the degrees of freedom compared to the Dirac fields \cite{Majorana,Dirac}.

In Majorana's theory, $\psi_c$ represents the charge conjugation of the spinor $\psi$ \cite{Majorana,Sanchez}. Therefore, if the additional condition $\psi=\psi_c$ is satisfied, the resulting spinor is known as a Majorana fermion. From a broader point of view, Majorana fermions can be seen as real solutions to the Dirac equation \cite{Sanchez,Zee}.

After recent studies on neutrino discovery, several researchers have turned their attention to the study of Majorana fermions \cite{Elliott}. Much of this interest is due to the fact that the physics of the Standard Model of elementary particles does not know whether neutrinos are their own antiparticles. Another scenario, for the application of the study of Majorana fermions, arises in the physics of condensed matter, e. g., in studies of topological superconductors \cite{Alicea}. Also, it is worth to mentioning that this theory has been used to describe theories applied to quantum computing as presented in Refs. \cite{Kitaev,Stanescu}. 

Indeed, among the Majorana fermions applications in the quantum computing scenario, consider that each qubit is a site that can be empty or occupied by an electron \cite{Kitaev}. Let us say that the particles that occupy this space are spin-up particles, so particles with another spin direction will be prohibited. Recent studies show that these sites are not exactly qubits, because in particular cases, the fermionic particles are Majorana fermions. One non-trivial observation is that it is theoretically possible to pair Majorana fermions by interaction, so that these fermions remain unpaired and separated from each other \cite{Elliott,Kitaev}. These systems can be used as decoherence-free quantum memory. It is a fact that a single Majorana fermion cannot interact with the environment on its own, since these particles are described by a non-physical operator. In this way, decoherence can only arise from an interaction mediated by the environment of two fermions of Majorana \cite{Elliott,Kitaev,Stanescu}.

In general, in recent years studies of quantum systems have attracted the attention of several researchers. In particular, there has been a growing interest in the study of simulations of relativistic quantum systems with Majorana equation in $(1+1)D$ \cite{Jin, Gerritsma}. An important fact is that studies using simulation methods in this quantum theory opens the possibility of implementing non-physical operations, such as complex conjugation, load conjugation and time inversion in several models \cite{Keil,Zhang}. 

On the other hand, in 1948, Claude E. Shannon built the concept that would be known as Shannon's entropy \cite{Shannon}. Although this concept originated in the mathematical theory of communication \cite{Shannon,Dong}, it has played an important role in the study of the quantum systems. The interpretation of Shannon's entropy in the space of positions is related to the uncertainty of the location of the particle in the space \cite{Dong1}. Likewise, entropy in the momentum space is related to the uncertainty of the momentum measurements of the particle. In this way, we can think that Shannon's entropy presents itself in quantum mechanics as a new formalism for the study of uncertainties related to quantum systems in non-relativistic and relativistic scenarios \cite{Shannon,Grosshans}.

In the last decades, many studies on the theory of quantum information, in particular, on Shannon's entropy, have been studied intensively by several researchers. We find in the current literature some theoretical studies on information measures for different quantum systems \cite{Solaimani,Shi11,Yahya}. Among these investigations, we have: study of systems with position-dependent mass \cite{Hua,Yanez11}, study of non-Hermitian models \cite{Lima}, Klein-Gordon oscillator \cite{Boumali}, study of the Schr\"odinger equation for several potentials, such as hyperbolic potential \cite{Valencia}, harmonic oscillator in D dimensions and hydrogen atom \cite{Dehesa}, etc.

Thermodynamic properties of quantum systems have also deserved attention for many years. Among the important works that treat this issue, we have studies of thermal properties of diatomic molecules in the presence of external magnetic fields \cite{Rampho}, study of the Denf-Fan-Eckart model \cite{Edet1}, study of the magnetic susceptibility of the Hellman potential in Aharanov-Bohm Flow with finite temperature \cite{Edet}, thermal study of the Poschl-Teller model \cite{Edet2}, thermal study in pseudo-harmonic potential \cite{Ikot11}, etc.

Throughout this work, we will turn our attention to the study of some statistical properties of linear Majorana fermions, namely, Shannon's entropy and thermodynamic properties. In particular, we believe that studying Majorana fermions can help us better understand the behavior of these relativistic particles. In fact, once we understand the behavior of these particles better, we are able to comprehensively understand some physical phenomena, e. g., the dynamic of  quantum spin liquid which has been widely studied \cite{Koga, Dusuel, Chaloupka, Jackeli} to understand spin transports in quantum spin systems and are applied in spintronic contexts \cite{LXIN}. In addition, to  understand the properties of Majorana's fermions allow us to know a little more about the thermodynamics associated with topological insulators and the Majorana's superconductivity \cite{Leijnse}, which can leads to semiconductor nanowires \cite{aasen} and to a revolution in the industry of electrical devices \cite{trif}.

The main objective of this work is to present, for the first time in the literature, a study of the influence of an external linear potential on Shannon's entropy and on the thermodynamic properties of the set of Majorana fermions. To achieve our objectives, we use the energy spectrum of the system and build the partition function in order to obtain the thermodynamic properties of the model, namely, free Helmholtz energy, entropy, mean energy and heat capacity. Therefore, we present, in an unprecedented way, the study of Shannon's entropy of particles that make up an ensemble of Majorana fermions. It is important to remark that for the study of Shannon's entropy we use the wave functions of the model. In other words, we consider only a single particle to investigate the position and moment uncertainties of the model subject to the influence of the linear potential. In contrast, when studying the thermodynamic properties, we consider the ensemble formed by these Majorana particles. 

This work is organized as follows. In Section II, we present a discussion of analytical solutions for linear Majorana fermions. In section III, we turn our attention to the study of Shannon's entropy and how this quantity changes with the parameter of the linear potential. In this way, we present the numerical and graphical results related to the Shannon entropy and the entropic densities of the model. Finally, in section IV, we build a canonical ensemble of these particles and study the thermodynamic properties of the system. Finally, we conclude by summarizing our results and making some discussions on the model.

\section{About linear Majorana fermions}

In this section, we studied linear Majorana fermions \cite{Sanchez}. In this case, considering the free Dirac equation, we have 
\begin{align}
    (i\gamma^{\mu}\hbar\partial_{\mu}-mc)\Psi_{D}=0,
\end{align}
where the $\gamma$-matrices are generators of Clifford algebra and obey the relation
\begin{align}
    \{\gamma^{\mu},\gamma^{\nu}\}=2\eta^{\mu\nu},
\end{align}
with the metric signature for the system in $(1+1)D$ being $\eta^{\mu\nu}=$ diag $(+,-)$. We remember that the $\gamma$-matrices are $2\times 2$ matrices for an irreducible representation. In this case, the Dirac $\Psi_D$ spinor is a function of two components that have two degrees of freedom associated with particles and their antiparticles. 

To study Majorana fermions, we must turn our attention to building an equation in which the fermionic particles are their own antiparticles. As a consequence, the Majorana spinor ($\Psi_M$) has half the degrees of freedom compared to the Dirac equation. We call attention to the fact that there is no spin in $(1+1)$ dimensions, therefore, strictly speaking, the Dirac equation in two dimensions does not describe half-integer spin particles. Mathematically, we should have the Majorana's representation $\Psi_{M}^{*}=\Psi_{M}$. 

The Majorana's representation is chosen such that 
\begin{align}
    \gamma^{0}=\begin{pmatrix}
0 & i \\
-i & 0 
\end{pmatrix}, \hspace{0.5cm} \gamma^{1}=\begin{pmatrix}
i & 0\\
0 & -i
\end{pmatrix}, \hspace{0.5cm} \text{and} \hspace{0.5cm} \gamma^{5}=i\gamma^{0}\gamma^{1}=\begin{pmatrix}
0 & i\\
i & 0
\end{pmatrix}.
\end{align}

In this way, we can write the Majorana equation in the presence of an external potential field as 
\begin{align}
\label{MajoranaE}
    \bigg[i\hbar\gamma^{\mu}\partial_{\mu}-mc\mathbf{I}-\frac{1}{c}\mathbf{I}\mathcal{V}\bigg]\Psi_{M}(x,t)=0,
\end{align}
where $\mathbf{I}$ is the identity matrix, $\mathbf{V}$ is the external potential field, $\hbar$ is the reduced Planck constant and $c$ the speed of light.

Interesting features arise when we study the case of linear Majorana fermions, i. e., assuming $\mathbf{V}=kx$. With that choice, we should have   
\begin{align}
\label{LinearM}
    \bigg[i\gamma^{\mu}\hbar\partial_{\mu}-mc-\frac{1}{c}kx\bigg]\Psi_{M}(x,t)=0.
\end{align}

The Majorana equation (\ref{MajoranaE}) can be rewritten in the form of the Schrödinger equation, i. e., 
\begin{align}
    i\hbar\frac{\partial}{\partial t}\Psi_M(x,t)=\mathcal{H}\Psi_{M}(x,t), 
\end{align}
with the Hamiltonian of the system being 
\begin{align}
    \mathcal{H}=c\alpha\hat{p}+\beta mc^2+\gamma^0\mathbf{V}.
\end{align}

Here we remark that $\mathcal{H}$ is the Dirac Hamiltonian operator, $\hat{p}$ is the momentum operator, $\beta=\gamma^0$ and $\alpha=\gamma^0\gamma^1$. 

We emphasize that Majorana fermions in the present model are neutral fermionic particles. With this choice of linear scalar potential, we observed that Majorana fermions are prevented from being coupled to another pseudo-scalar potential. 

Rewriting Majorana's equation in terms of its components, i. e., $\Psi_M=(\psi_1,\psi_2)^{T}$, we get the coupled equations, namely, 

\begin{align}
\label{M1}
    \hbar\partial_t \psi_1 (x,t)=[-c\hbar\partial_x mc^2+kx]\psi_2 (x,t),
\end{align}
and
\begin{align}
\label{M2}
    -\hbar\partial_t \psi_2 (x,t)=[c\hbar\partial_x+mc^2+kx]\psi_1 (x,t).
\end{align}

Using the factorization method, which is well known from supersymmetric quantum mechanics scenarios, we built two supersymmetric operators to solve the differential equation. Thus, we define: 
\begin{align}
    \hat{A}=c\hbar\partial_x+mc^2+kx, \hspace{0.5cm} \text{and} \hspace{0.5cm} \hat{A}^{\dagger}=c\hbar\partial_x+mc^2+kx.
\end{align}

Considering the supersymmetric operators, we decouple the Majorana equation and rewrite the eqs. (\ref{M1}) and (\ref{M2}), as follows, 
\begin{align}
\label{ME1}
    -\hbar^2\partial_{t}^{2}\psi_1=\hat{A}^{\dagger}\hat{A}\psi_1,
\end{align}
and
\begin{align}
\label{ME2}
    -\hbar^2\partial_{t}^{2}\psi_2=\hat{A}\hat{A}^{\dagger}\psi_2.
\end{align}

Here, we define that the partners Hamiltonian of the model are: 
\begin{align}
    \hat{H}_{\pm}=-c^2\hbar^2\frac{d^2}{dx^{2}}+V_{\pm}.
\end{align}

Therefore, the supersymmetric Majorana eqs. (\ref{ME1}) and (\ref{ME2}) are reduced to
\begin{align}
    \hbar^2\partial_{t}^{2}\psi_1=\hat{H}_{-}\psi_1,
\end{align}
and
\begin{align}
    \hbar^2\partial_{t}^{2}\psi_2=\hat{H}_{+}\psi_2,
\end{align}
where
\begin{align}
    \hat{H}_{\pm}=-c^2\hbar^2\frac{d^2}{dx^{2}}+(mc^{2}+kx)^2+\pm c\hbar k.
\end{align}

With the hypothesis that $\hat{H}_{\pm}$ are time independent and the components $\psi_1$ and $\psi_2$ are functions of real values, we use variable separation and consider the ansatz 
\begin{align}
    \psi_{1,2}(x,t)=\phi_{\mp}(x)T_{\mp}(t).
\end{align}

In this way, we find solutions in the form
\begin{align}
    \psi_1 (x,t)=\phi_{-}(x)\sin\bigg[\frac{\mathcal{E}t}{\hbar}+\Omega\bigg],
\end{align}
and
\begin{align}
    \psi_2 (x,t)=\phi_{+}(x)\cos\bigg[\frac{\mathcal{E}t}{\hbar}+\Omega\bigg],
\end{align}
where the parameter $\Omega$ defines the behavior at the initial time of the functions $\psi_{1,2}$. The functions $\phi_{\pm}$ obey the equations 
\begin{align}
\label{Schr}
    \hat{H}^{\pm}\phi_{\pm}(x)=\mathcal{E}^{2}\phi_{\pm}(x).
\end{align}

Each state with $\mathcal{E}\neq 0$ has a one-to-one mapping between the energy states of $\phi_{\pm}$, i. e., 
\begin{align}
    \hat{A}^{\dagger}\phi_{+}=\frac{1}{\mathcal{E}}\phi_{-} \hspace{0.5cm} \text{and} \hspace{0.5cm} \hat{A}\phi_{-}=\frac{1}{\mathcal{E}}\phi_{+}.
\end{align}

Note that solving Majorana's equation in the presence of an external static and linear field is reduced to solving a supersymmetric quantum mechanics problem. Here, we also call attention to the fact that when one of the Hamiltonians has an energy state $\mathcal{E}=0$, the SUSY model is called unbroken. Otherwise, SUSY is called broken \cite{Sanchez}. In the unbroken case, one of the Hamiltonians has an additional zero-energy self-state that does not appear in his partner Hamiltonian. The zero energy self-states of $ \hat{H}_{+}(\hat{H}_{-})$ can be determined by imposing that it is annihilated by the operator $\hat{A}^{\dagger}(\hat{A})$. 

If we assume that $\hat{A}\phi_{0_{-}}=0$ and $\phi_{0_{-}} $ is normalizable, we have $\mathcal{E}_{0}^{-}=0$, and consequently 
\begin{align}
    \mathcal{E}_{n}^{+}=\mathcal{E}_{n+1}^{-},
\end{align}
with wave functions respecting equality,
\begin{align}
\label{rel1}
    \phi_{n_{+}}=(\mathcal{E}_{n+1}^{-})^{-1}\hat{A}\phi_{(n+1)_{-}}, \hspace{0.5cm} \text{with} \hspace{0.5cm} n=0,1,2,3,...,
\end{align}
meanwhile, 
\begin{align}
    \phi_{(n+1)_{-}}=(\mathcal{E}_{n}^{+})^{-1}\hat{A}^{\dagger}\phi_{n_{+}}^{\dagger} \hspace{0.5cm} \text{with} \hspace{0.5cm} n=1,2,3,...
\end{align}

A similar condition can be adopted if we have $\hat{A}^{\dagger}\phi_{+}^{0}=0$ and $\phi_{+}$ normalizable. In this case, the relations are:
\begin{align}
        &\mathcal{E}_{n}^{+}=\mathcal{E}_{n+1}^{-}, \hspace{0.5cm} \mathcal{E}_{0}^{+}=0,\\
        &\phi_{n_{-}}=(\mathcal{E}_{n+1}^{+})^{-1}\hat{A}\phi^{\dagger}_{(n+1)_{+}}, \hspace{0.5cm} \text{with} \hspace{0.5cm} n=0,1,2,3,...,\\
        &\phi_{(n+1)_{+}}=(\mathcal{E}_{n_{-}})^{-1}\hat{A}\phi_{n_{-}}.
\end{align}

Note that the solutions of the Majorana equation do not result in stationary self-states, as can be verified by calculating the probability density. However, it is observed that the ground state is an exception. 

We observed that if the supersymmetric Hamiltonian depends on a real parameter in such way that it is invariant under a discrete reparametrization, it is possible to obtain the general algebraic spectrum. Thus, if $\hat{H}_{-}$ has a zero eigenvalue, the invariance condition is given by \begin{align}
    \hat{H}_{+}(\xi_1,x)=\hat{H}_{-}(\xi_2, x)+\eta(\xi_1).
\end{align}

The partners Hamiltonians have the same shape differing by a constant $\eta$ which depends on the parameter $\xi_j$. The parameter $\xi_2$ is a function of $\xi_1$, i. e., $\xi_2=f(\xi_1)$. Through successive reparametrization and using the profile invariance condition, we have 
\begin{align}
    \hat{H}_{+}(a_{l-1},x)=\hat{H}_{-}(a_l,x)+R(a_{l-1}).
\end{align}

From this we have that the energy spectrum of eq. (\ref{Schr}) is 
\begin{align}
\label{Energy0}
    \mathcal{E}_{n}^{-}(\xi_1)=\pm\sqrt{\sum_{k=1}^{n}R(\xi_k)}, \hspace{0.5cm} \text{and} \hspace{0.5cm} \mathcal{E}_{0}^{-}(\xi_1)=0.
\end{align}

To obey the profile invariance condition, we assume that $\xi_l=k$ and $R(\xi_l)=2c\hbar k$. Hence, 
\begin{align}
\label{Energy00}
    \mathcal{E}_{n}^{-}=\pm\sqrt{\sum_{k=1}^{n}2c\hbar k}=\pm\sqrt{2c\hbar kn},
\end{align}
with $\mathcal{E}_{0}^{-}=0$.

In this case, $\hat{A}\phi_{0_{-}}=0$ and the ground state is given by
\begin{align}
\label{ground}
    \phi_{0_{-}}(y)=\bigg(\frac{\omega}{\pi}\bigg)^{1/4}\exp\bigg(-\frac{\omega}{2}y^2\bigg),
\end{align}
with $\omega=k/c\hbar$ and $y=x+mc^2/k$.

Note that $k>0$, so the wave function is normalized. To obtain the other states, we must make successive applications of $\hat{A}^{\dagger}$ in eq. (\ref{ground}). From this, we have 
\begin{align}
\label{sol}
    \phi_{n_{-}}(y)=\frac{1}{2^{\frac{n}{2}}(n!)^{\frac{1}{2}}}\bigg(\frac{\omega}{\pi}\bigg)^{1/4}\exp\bigg(-\frac{\omega}{2}y^2\bigg)H_{n}(\omega^{\frac{1}{2}}y),
\end{align}
where $H_n$ are the Hermite polynomials.

Substituting the result presented in eqs. (\ref {sol}) and  (\ref{rel1}). We finally get that 
\begin{align}
    \phi_{n_{+}}(y)=\frac{1}{2^{\frac{n-1}{2}}[(n-1)!]^{\frac{1}{2}}}\bigg(\frac{\omega}{\pi}\bigg)^{1/4}\exp\bigg(-\frac{\omega}{2}y^2\bigg)H_{n-1}(\omega^{\frac{1}{2}}y).
\end{align}

In this way, we have that the complete solution for linear Majorana fermions is 
\begin{align}
\label{functionY}
    \psi_n(y,t)=\frac{\omega^{\frac{1}{4}}\exp(-\omega y^2/2)}{2^{\frac{n}{2}}(n!)^{\frac{1}{2}}\pi^{\frac{1}{4}}}\begin{pmatrix}H_{n}(\omega^{\frac{1}{2}}y)\sin(\sqrt{2\omega n}ct+\Omega)\\
    \sqrt{2n}H_{n-1}(\omega^{\frac{1}{2}}y)\cos(\sqrt{2\omega n}ct+\Omega)
    \end{pmatrix}, \hspace{0.5cm} \text{with} \hspace{0.5cm} n=1,2,3,...
\end{align}
and for $n=0$, we have

\begin{align}
    \label{functionY1}
    \psi_0(y,t)=\frac{\omega^{\frac{1}{4}}\exp(-\omega y^2/2)}{\pi^{1/4}}\begin{pmatrix}
    1\\ 0
    \end{pmatrix}.
\end{align}

\begin{figure}
\begin{center}
\begin{tabular}{ccc}
\includegraphics[scale=0.3]{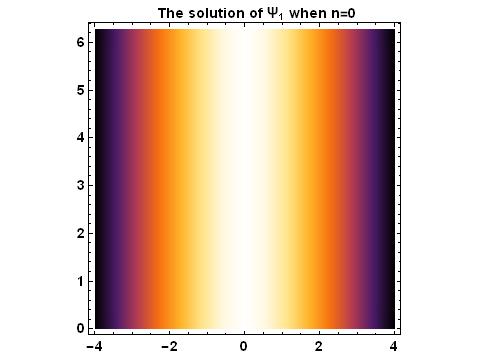}\\
\includegraphics[scale=0.35]{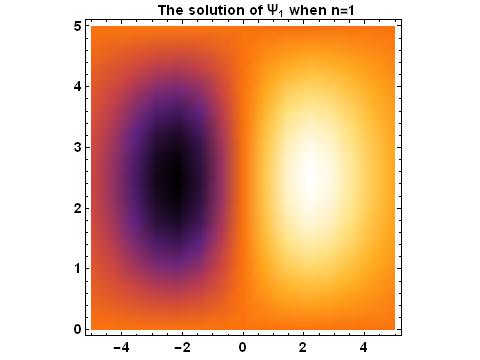}
\includegraphics[scale=0.35]{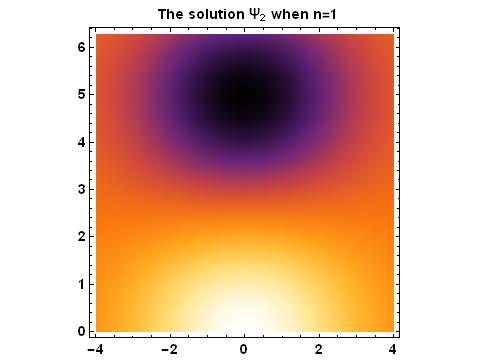}\\
\includegraphics[scale=0.35]{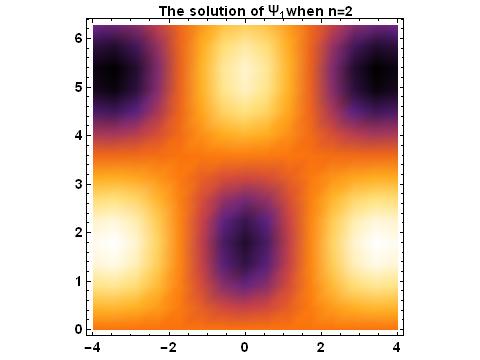}
\includegraphics[scale=0.35]{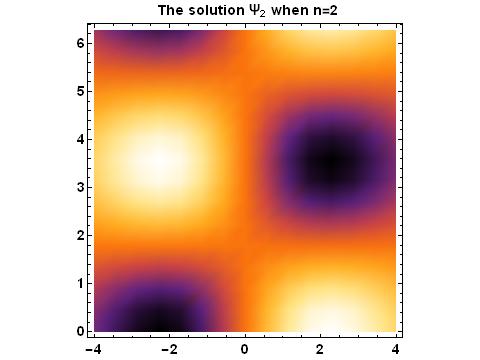}\\ 
\includegraphics[scale=0.35]{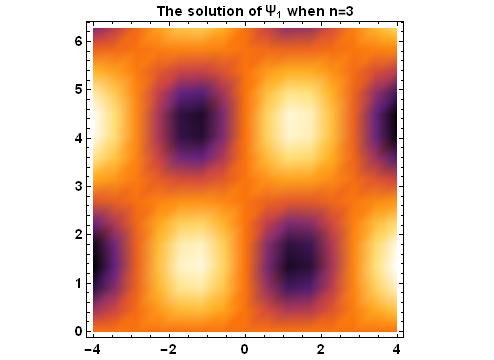}
\includegraphics[scale=0.35]{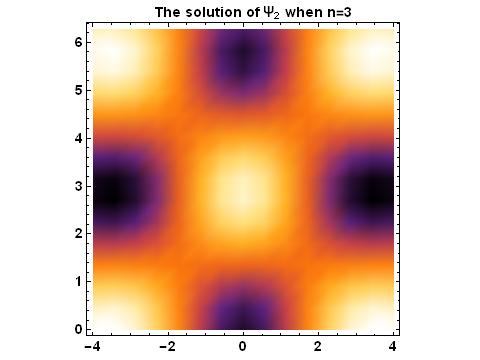}\\ 
\end{tabular}
\end{center}
\caption{Planar solution of Majorana fermions (with  $\omega=0.2$) evolving over time. The light points represent the maximal critical point of the wave function in an instant of time $t$.
\label{figure0}}
\end{figure}

We show in Fig. (\ref{figure0}) the planar behavior of linear Majorana fermions evolving over time. From figure (\ref{figure0}) it is clear the existence of forbidden regions (dark points) for the linear Majorana fermions and regions of the most probable space of localization of the particles (light points). It is also clear that these regions are shifting as time progresses.  

\section{Shannon's entropy}

Let us to observe that the study of Shannon's entropy in Majorana fermions allows us to have advances in physics. For instance, with the analysis of this quantity in our model it is possible to better understand how the intensity of a potential modifies the uncertainties of the probabilities of these particles. In other words, it is possible to notice how an external force influences the measured uncertainties of the Majorana particle. Not far from this point of view, in the scenario of communication theory we would have information on how these particles can ``propagate information in the best possible way''. It is evident that, depending on our objective, Shannon's entropy can give us a different interpretation of the problem. We will discuss in this work, that Shannon's entropy allows us to understand more clearly the uncertainties associated with linear Majorana fermions. It is interesting to mention that our work, as far as we know, is pioneer in the study of the uncertainties associated with these particles. We believe that understanding these systems and their measurement uncertainties can allow us to have significant advances in the study of phenomena such as superconductivity and in the study of topological insulators. Here it is important to mention that these phenomena are of wide interest, going from physics to chemistry and engineering.

From this point on, we will turn our attention to the study of Shannon's entropy of the model. We know that entropic uncertainty relation is an alternative to Heisenberg's uncertainty principle in quantum mechanics \cite{Aptekarevh}. The concept of entropic uncertainty provides the irreversibility of the physical system in thermodynamic. According to statistical physics, we know that the entropy measure appears as a measure associated with the degree of disorder in the system. Among the various information measures, Shannon entropy \cite{Shannon} plays a particularly important role in measuring uncertainty. Shannon's entropy for the nth-state in the space of the position is defined by 
\begin{eqnarray}\label{0.11}
S_{y}^{n}=-\int_{-\infty}^{\infty}|\Psi_n(y,t)|^{2}\ln|\Psi_n(y,t)|^{2}dy,
\end{eqnarray}
and, in the reciprocal space is 
\begin{eqnarray}\label{0.12}
S_{p}^{n}=-\int_{-\infty}^{\infty}|\Psi_n(p,t)|^{2}\ln|\Psi_n(p,t)|^{2}dp,
\end{eqnarray}
where $\rho_n=|\Psi_n|^2$ is the probability density, and $\Psi_n(p,t)$ is the wave function in the momentum space.

Shannon's entropy must obey the entropic uncertainty relation. The BBM relation between the position space and momentum space was obtained by Beckner, Bialynicki-Birula and Mycielski (BBM) \cite{BBM}. This relation is written as 
\begin{equation}
S_{y}^{n}+S_{k}^{n}\geq D(1+\ln\pi),
\end{equation}
where $D$ is the spacial dimension of the model.

Now, we carry out an entropic analysis on linear Majorana fermions. We obtain the solutions of the function that describes the fermionic particles as given by Eq. (\Ref{functionY}). Thus, without losing generality, we analyze the first energy levels of the model, i. e., $ n = 0,1, 2, $ and $ 3 $, namely:
\begin{eqnarray}
\label{1}
\Psi_0(y,t)&=&\frac{\omega^{\frac{1}{4}e^{-\frac{\omega}{2}y^2}}}{\pi^{\frac{1}{4}}}\left(\begin{array}{cccccc}
1\\
0\\
\end{array}\right),\\
\Psi_1(y,t)&=&\frac{\omega^{\frac{1}{4}e^{-\frac{\omega}{2}y^2}}}{\sqrt{2}\pi^{\frac{1}{4}}}\left(\begin{array}{cccccc}
(2\omega)^{\frac{1}{2}}y \sin[c(2\omega)^{\frac{1}{2}}t+\delta]\\
\cos[c(2\omega)^{\frac{1}{2}}t+\delta]\\
\end{array}\right),
\\
\Psi_2(y,t)&=&\frac{\omega^{\frac{1}{4}e^{-\frac{\omega}{2}y^2}}}{2\pi^{\frac{1}{4}}}\left(\begin{array}{cccccc}
(2\omega y^2-1) \sin[c(4\omega)^{\frac{1}{2}}t+\delta]\\
2\omega^{\frac{1}{2}} y \cos[c(4\omega)^{\frac{1}{2}}t+\delta]\\
\end{array}\right),
\\
\Psi_3(y,t)&=&\frac{\omega^{\frac{1}{4}e^{-\frac{\omega}{2}y^2}}}{2\sqrt{6}\pi^{\frac{1}{4}}}\left(\begin{array}{cccccc}
2\omega^{\frac{1}{2}} y(2\omega y^2-3) \sin[c(6\omega)^{\frac{1}{2}}t+\delta]\\
\sqrt{6} (2\omega y^2-1)\cos[c(6\omega)^{\frac{1}{2}}t+\delta]\\
\end{array}\right).
\end{eqnarray}

\begin{figure}[ht!]
\begin{center}
\begin{tabular}{ccc}
\includegraphics[scale=0.55]{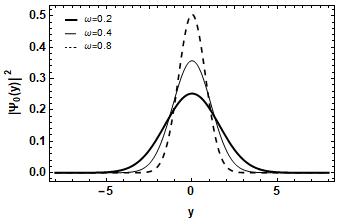}
\includegraphics[scale=0.55]{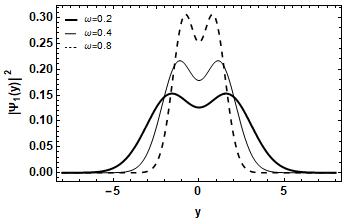}\\ 
(a) \hspace{8 cm}(b)\\
\includegraphics[scale=0.55]{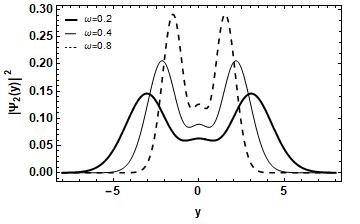}
\includegraphics[scale=0.55]{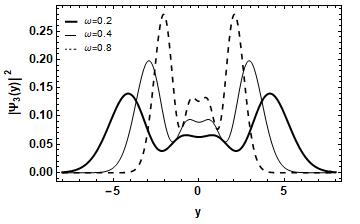}\\
(c) \hspace{8 cm}(d)
\end{tabular}
\end{center}
\caption{Plots of the probability density: (a) $n=0$, (b) $n=1$, (c) $n=2$ and (d) $n=3$.
\label{fig1}}
\end{figure}

In Figure (\ref{fig1}), we plot the propability densities, i. e., $|\Psi_n(y, t)|^2$. We can notice that as the energy levels rise, new critical points of probability are clearly seen in Fig. (\ref{fig1}b) when $n=2$ and in Fig. (\ref{fig1}c) when $n=3$.

We can find the wave functions in reciprocal space through the Fourier transform, namely, 
\begin{equation}
\Psi_n(p,t)=\frac{1}{\sqrt{2\pi}}\int_{-\infty}^{\infty}\Psi_n(y,t)\,\text{e}^{ipy} dy.
\end{equation}

Therefore, for the first energy levels, i. e., $n=0,1,2,3$, we have: 
\begin{eqnarray}
\label{2}
\Psi_0(p,t)&=&\frac{e^{-\frac{p^2}{2\omega}}}{(\omega\pi)^{\frac{1}{4}}}\left(\begin{array}{cccccc}
1\\
0\\
\end{array}\right),\\
\Psi_1(p,t)&=&\frac{e^{-\frac{p^2}{2\omega}}}{\sqrt{2}(\omega^3\pi)^{\frac{1}{4}}}\left(\begin{array}{cccccc}
i\sqrt{2}p \sin[c(4\omega)^{\frac{1}{2}}t+\delta]\\
\omega^{\frac{1}{2}} \cos[c(4\omega)^{\frac{1}{2}}t+\delta]\\
\end{array}\right),
\\
\Psi_2(p,t)&=&\frac{e^{-\frac{p^2}{2\omega}}}{2(\omega^5\pi)^{\frac{1}{4}}}\left(\begin{array}{cccccc}
(\omega-2p^2) \sin[c(4\omega)^{\frac{1}{2}}t+\delta]\\
2i\omega^{\frac{1}{2}}p \cos[c(4\omega)^{\frac{1}{2}}t+\delta]\\
\end{array}\right),
\\
\Psi_3(p,t)&=&\frac{e^{-\frac{p^2}{2\omega}}}{2\sqrt{6}(\omega^7\pi)^{\frac{1}{4}}}\left(\begin{array}{cccccc}
-ip(4p^2-6\omega) \sin[c(6\omega)^{\frac{1}{2}}t+\delta]\\
\sqrt{6}\omega^{\frac{1}{2}}(\omega-2p^2) \cos[c(6\omega)^{\frac{1}{2}}t+\delta]\\
\end{array}\right).
\end{eqnarray}

\begin{figure}[ht!]
\begin{center}
\begin{tabular}{ccc}
\includegraphics[scale=0.55]{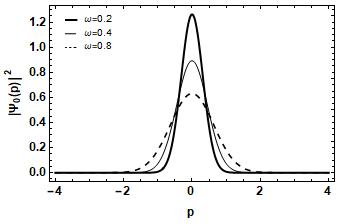}
\includegraphics[scale=0.55]{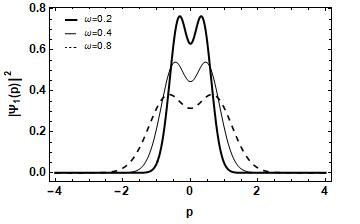}\\ 
(a) \hspace{8 cm}(b)\\
\includegraphics[scale=0.55]{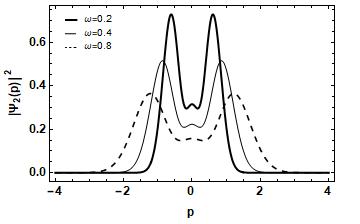}
\includegraphics[scale=0.55]{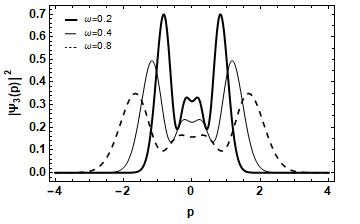}\\
(c) \hspace{8 cm}(d)
\end{tabular}
\end{center}
\caption{Plots of the probability density in the momentum space: (a) $n=0$, (b) $n=1$ and (c) $n=2$ and (d) 
 $n=3$.
\label{fig2}}
\end{figure}

The behavior of the probability density for the momentum space $ | \ Psi_n (p, t) | ^ 2 $ is shown in Fig. (\ref {fig2}). As expected, the probability density becomes more localized in the momentum space. We also noticed the appearance of new critical points when the energy levels increase. 

\begin{figure}
\begin{center}
\begin{tabular}{ccc}
\includegraphics[scale=0.5]{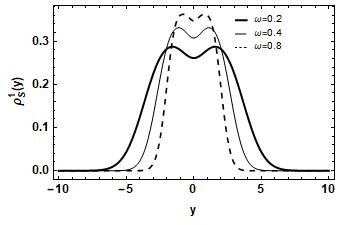}\\ 
(a)\\
\includegraphics[scale=0.5]{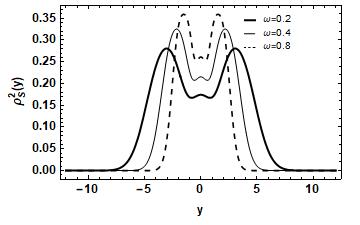}
\includegraphics[scale=0.5]{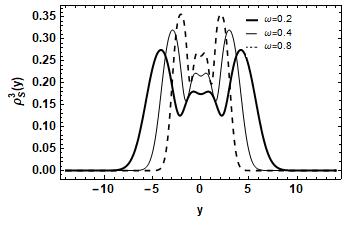}\\
(b) \hspace{8 cm}(c)
\end{tabular}
\end{center}
\vspace{-10pt}
\caption{Plots of the entropic density in the position space: (a) $n=1$, (b) $n=2$ and (c) $n=3$.
\label{fig3}}
\end{figure}

Through Shannon's entropy demonstrated in the equations (\ref{0.11}) and (\ref{0.12}), we can define the entropic density as: 
\begin{eqnarray}
\rho^n_{S}(y)=|\varphi_n(y)|^2\ln|\varphi_n(y)|^2,
\end{eqnarray}
and
\begin{eqnarray}
\rho^n_{S}(p)=|\varphi_n(p)|^2\ln|\varphi_n(p)|^2.
\end{eqnarray}
Since they play a role similar to the probability density $\rho_n=|\Psi_n|^2$, in Fig. (\ref{fig3}), we present the illustration of the behavior of the entropic density of the studied model. We noticed a behavior similar to that observed in Fig. (\ref{fig1}), where by increasing the energy levels we obtain additional critical points. In Fig. (\ref{fig3}a), we notice that by increasing the value of the parameter $\omega$ we obtain new points in entropic density and a more localized entropic density. The same happens in Figs. (\ref{fig3}b) and (\ref{fig3}c).

\begin{figure}
\begin{center}
\begin{tabular}{ccc}
\includegraphics[scale=0.5]{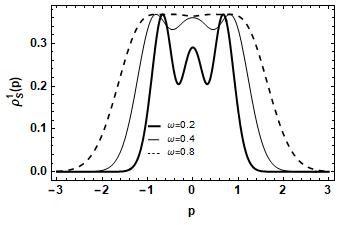}\\ 
(a)\\
\includegraphics[scale=0.5]{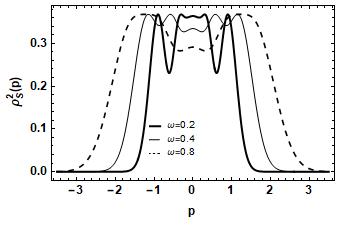}
\includegraphics[scale=0.5]{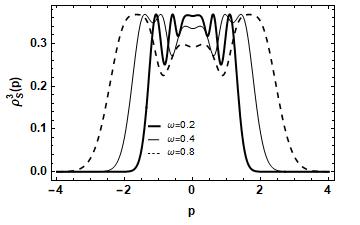}\\
(b) \hspace{8 cm}(c)
\end{tabular}
\end{center}
\vspace{-10pt}
\caption{Plots of the entropic density in the reciprocal space: (a) $n=1$, (b) $n=2$ and (c) $n=3$.
\label{fig4}}
\end{figure}

In the momentum space, the entropic density have a behavior contrary to the position space one, where decreasing the value of $\omega$ new points become notables, as can be seen in the Fig. (\ref{fig4}). We also observed that by increasing the energy level, new critical points in entropic density emerge, a behavior similar to the position space one.

\begin{figure}[ht!]
\begin{center}
\begin{tabular}{ccc}
\includegraphics[scale=0.55]{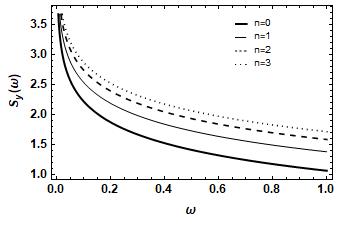}
\includegraphics[scale=0.55]{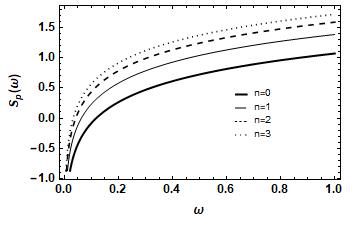}\\
(a) \hspace{8 cm}(b)
\end{tabular}
\end{center}
\caption{Shannon's entropy for linear Majorana fermions: $S_y(\omega)$(a)  and $S_p(\omega)$(b).} \label{fig5}
\end{figure}

With the solutions in the space of positions (\ref{1}) and in the space of momenta (\ref{2}), we can make an analysis of the numerical results for the Shannon entropy presented in the Table (\ref{tab1}). The Fig. (\ref{fig5}) helps us to better understand the entropy behavior by varying the parameter $\omega$. We note that by increasing the $\omega$ parameter, Shannon's entropy tends to decrease in the position space and increase in the momentum space, i. e., when we increase the parameter that adjusts the potential (or the force acting on the particle) of Majorana fermions, Shannon's entropy decreases in the space of the positions and increases in the space of the momentum in such a way that the BBM relation is satisfied. We also observed that when we increase energy levels, Shannon's entropy tends to increase in both spaces, therefore increasing the uncertainty of measurements in these spaces. 

\begin{table}[h!]\label{tab1}
\centering
\begin{tabular}{|c|c|c|c|c|c|}\hline
\hline
$n$ & $\omega$ & $S_{y}$ & $S_{p}$ & $S_{y}+S_{p}$ & $1+\ln\pi$\\ \hline
\hline
  & 0.2   & 1.87708  & 0.26765  & 2.14473& \\
0 & 0.4  & 1.53051 & 0.61422  & 2.14473& 2.14473 \\
  & 0.8 & 1.18394 & 0.96079  & 2.14473 &\\ \hline
  & 0.2   & 2.19246  & 0.58302  & 2.77548& \\
1 & 0.4  & 1.84588 & 0.92959  & 2.77548& 2.14473 \\
  & 0.8 & 1.49931 & 1.27617  & 2.77548 &\\ \hline
  & 0.2   & 2.39707  & 0.78763  & 3.18469& \\
2 & 0.4  & 2.05049 & 1.13420  & 3.18469& 2.14473\\
  & 0.8 & 1.70392 & 1.48078  & 3.18469& \\ \hline
  & 0.2   & 2.52764  & 0.91820  & 3.44584& \\
3 & 0.4  & 2.18107 & 1.26477  & 3.44584 & 2.14473\\
  & 0.8 & 1.83449 & 1.61135  & 3.44584 & \\ \hline
\end{tabular}\\
\caption{Numerical result of the Shannon's entropy for linear Majorana fermions.}
\label{tab1}
\end{table}

\section{Thermodynamic properties}

The main motivation for our study of the thermodynamic properties of an ensemble of Majorana fermionic particles, is due to the fact that by understanding the properties of these particles we can better understand how they can interfere in the quantum computing scenario. Since we know that in this scenario, quantum memory is achieved through the pairing of an ensemble of fermionic particles that, in principle, may be the well-known Majorana fermions. Indeed, when we study the thermodynamic properties of such particles, we are eventually making an attempt to collaborate with this theory. In other words, if we think that quantum memory is made up of this ensemble of fermionic particles, then we have that these particles are associated with a thermodynamic description. In this way, our study can contribute significantly to advances in several areas, namely, theoretical and experimental physics, engineering, etc.

There are several works in the current literature that discuss the thermodynamic properties of relativistic \cite{Oliveira} and non-relativistic \cite{Rampho,Edet1,Edet,Ikot11} systems. Motivated by some of these contributions, we will now turn our attention to the study of the model's thermodynamic properties. To achieve our goal, we start by defining the partition function of an ensemble of fermionic particles in a thermal bath given by 
\begin{equation}
\label{partition0}
   Z=\sum_{n=0}^{\infty}\text{e}^{-\beta\mathcal{E}_{n}}, 
\end{equation}
where $\beta=1/k_{B}T$; $k_{B}$ is the Boltzmann constant and $T$ is the temperature of the ensemble in thermodynamic equilibrium.

It is important to remember that the partition function plays a fundamental role in statistical mechanics, and that all thermodynamic quantities are built from it \cite{Lima}. After obtaining the $Z$ partition function, all properties of the linear Majorana fermions can be investigated. The main thermodynamic functions of our interest are the Helmholtz free energy $F$, the mean energy $U$, the entropy $S$ and the heat capacity $C_V$ which are defined as 
\begin{align}
 F=-\frac{1}{\beta}\text{ln}\, Z_N; \hspace{1cm} U=-\frac{\partial}{\partial\beta}\text{ln}\, Z_N; \hspace{1cm} S=k_{B}\beta^{2}\frac{\partial\mathcal{F}}{\partial \beta}; \hspace{1cm} C_V=-k_{B}\beta^{2}\frac{\partial U}{\partial\beta},
\end{align}
where $Z_N$ is the total partition function for the ensemble of linear Majorana N-fermions. 

Let us build the partition function for Majorana fermions. For this, we consider that the energy of the fermions is described by Eq. (\ref{Energy0}). In this way, the partition function is 
\begin{align}
\label{partition1}
    Z=\sum_{n=0}^{\infty}\text{e}^{\mp\beta\sqrt{2c\hbar kn}}.
\end{align}

To analyze the summation above, we consider the Euler-Maclaurin formula and assume the weak interaction limit, i. e., $ k \ll 1 $. Thus, we conclude that 
\begin{align}
\label{partition2}
    Z=\sum_{n=0}^{\infty}\text{e}^{-\beta\mathcal{E}_{n}}\cong\frac{1}{2}f(0)+\int_{0}^{\infty}\text{e}^{\mp\beta\sqrt{2c\hbar kx}}dx-\sum_{p=1}^{\infty}\frac{1}{(2p)!}B_{2p}f^{2p-1}(0).
\end{align}
Here we remember that $B_{2p}$ are Bernoulli's numbers. 

From expression (\ref{partition2}), and considering fermionic particles with energy $\mathcal{E}_n=\sqrt{2c\hbar kn}$, we will have  
\begin{align}
\label{partition3}
Z\simeq\frac{1}{2}+\frac{1}{c\hbar k\beta^2}.
\end{align}

For the indistinguishable $N$-fermions of Majorana, we obtain that
\begin{align}
\label{partition4}
    Z_N=\bigg[\frac{1}{2}+\frac{1}{c\hbar k\beta^2}\bigg]^N.
\end{align}

Therefore, using the partition function (\ref{partition4}), the Helmholtz free energy, the mean energy, the entropy and the heat capacity for the case studied can be written as follows: 
\begin{align}
    \label{thermodynamic}
    &F\simeq -\frac{N}{\beta}\text{ln}\bigg[\frac{1}{2}+\frac{1}{c\hbar k\beta^2}\bigg], \hspace{0.5cm} U\simeq\frac{4}{\beta(2+c\hbar k\beta^2)},\\
    &S\simeq  \frac{4Nk_B}{2+c\hbar k\beta^2}+Nk_B\text{ln}\bigg[\frac{1}{2}+\frac{1}{c\hbar k\beta^2}\bigg], \\
    &C_V\simeq \frac{4k_B N(2+3c\hbar k\beta^2)}{(2+c\hbar k\beta^2)^2}.
\end{align}

\begin{figure}
\begin{center}
\begin{tabular}{ccc}
\includegraphics[scale=0.4]{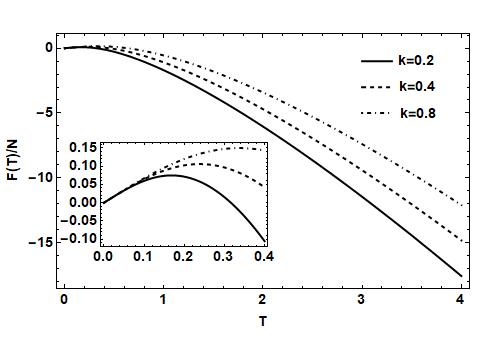}
\includegraphics[scale=0.4]{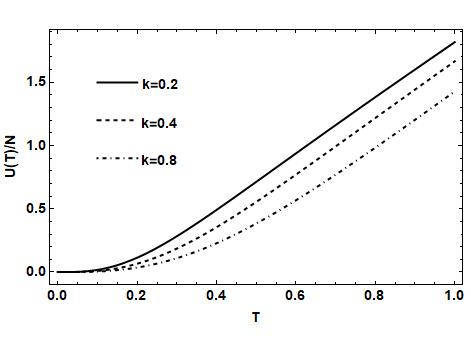}\\ 
(a) \hspace{7cm} (b)\\

\includegraphics[scale=0.4]{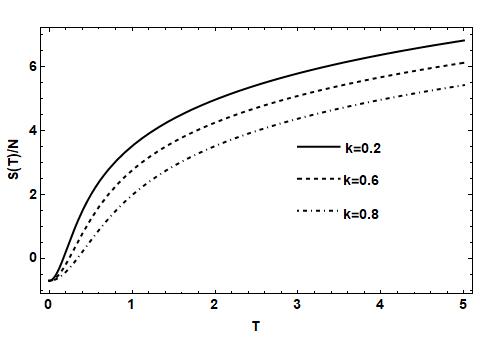}
\includegraphics[scale=0.4]{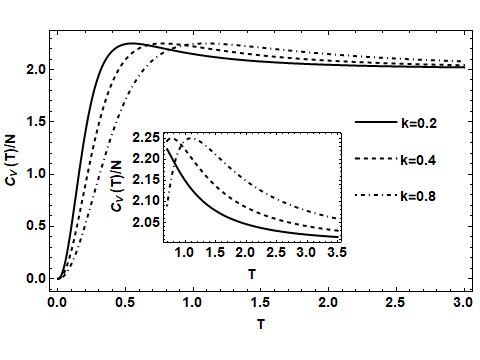}
\\
(c) \hspace{7cm}(d)
\end{tabular}
\end{center}
\caption{Thermodynamic properties of linear Majorana fermions for several values of the $k$ parameter.
\label{fig6}}
\end{figure}

The graphs for the thermodynamic functions of the linear Majorana fermions are provided in Fig. (\ref{fig6}). We consider $c=\hbar=k_B=1$. 

It is important to note that for the case of an ensemble of fermionic anti-particles at the weak interaction limit, the thermodynamic properties of linear Majorana fermions are identical. Therefore, Fig. (\ref{fig6}) represents the thermodynamic properties of particles as well as of anti-particles. 

\section{Results and discussions}

Through the numerical results presented in Table 1 and the graphical results presented in Figs. $(3)$, $(4)$, and $(5)$, we notice that Shannon's entropy tends to decrease in the space of the positions as the parameter $k$ increases its contribution. In other words, when the linear force increases the intensity in the linear Majorana fermions, Shannon's entropy decreases. When analyzing the result in the space of momenta, we notice that as the parameter $k$ increases, Shannon's entropy in the reciprocal space decreases respecting the entropic uncertainty relation. We also noticed that the more intense the force on linear Majorana fermions, the more localized is Shannon's entropic density in the space of the positions. In this way, we can conclude that the uncertainties in the measurement of the location of Majorana fermions will be less.

In contrast, in reciprocal space, we have that entropic densities are more localized when the force is less intense, i.e., for lower values of $k$. This fact leads us to conclude that the uncertainties in the measurements related to the particle momentum are more significant. 

With the energy spectrum found in the expression (\ref{Energy00}) by some statistical arguments we consider an ensemble of Majorana fermions in a thermal bath partition function and study the influence of force, i. e., of the parameter $k$ in the statistical properties, namely, Helmholtz free energy, mean energy, entropy, heat capacity. Remember that in all calculations we assume that $c=\hbar=k_{B}=1$.

In Fig. (6), we trace all profiles of thermal quantities as a function of temperature $T$ for different values of the parameter $k$ responsible for the contribution in the intensity of the strength of the fermionic particles, namely, $k=0.2,0.4$ and $0.8$. Due to the thermodynamic properties, we note that the Helmholtz free energy function $F(T)/N$ decreases with temperature almost exponentially. When the $T$ increases to higher values of the parameter $k$, it is concluded that the contribution of Helmholtz free energy becomes more significant in the ensemble of fermionic particles. In other words, Helmholtz free energy tells us that for a more intense force a system will have a greater amount of energy to realize work.

We also observed that when the parameter $k$ increases, the mean energy $U/N$ increases with an approximately linear behavior close to the origin, i. e., $T=0$. However, when $T$  increases, the behavior of the  energy starts to be more distinct. We also observed that for more intense forces, in the regime of high temperature the energy will assume an exponential behavior. Thus, we conclude that for more intense force the system has the capacity to increase the transfer of matter, or energy in the form of heat at the high temperature limit. Finally, we realize that for various values of $k$, the thermal capacity of the system tends to a constant, i. e., $C_V(T\rightarrow\infty)/ N\rightarrow 2k_{B}$ in this relativistic case. 

\section{Conclusion}
In this work, we investigate Shannon's entropy and the thermodynamic properties of linear Majorana fermions in a thermal bath in the canonical ensemble at finite temperature. We call linear Majorana fermions those that submitted to a 1D linear potential as $V=kx$.  For Shannon's entropy, we calculate the wave functions. With this result, we note that Shannon's entropy of Majorana fermions satisfies the BBM relation for fundamental states and excited states independent of the value of the parameter $k$. We conclude that the parameter $k$ has the capacity to modify the entropy of Shannon, i. e., an external force that acts on the Majorana fermions alters Shannon's entropy. However, this alteration obeys the BBM relation of the model.

In order to obtain thermodynamic properties, we calculate the energy spectrum for the case studied. We realized that the exact partition function did not have a closed form to perform our calculations, so we assumed the weak interaction approach and used the Euler-Maclaurin sum formula to evaluate the partition function numerically. Once the partition function was determined, all the main thermodynamic quantities could be derived, that is, the free energy of Helmholtz $F$, the mean energy $U$, the entropy $S$ and the capacity heat $C_V$ and its respective graphs were plotted for several values of $T$ and different values of the parameter $k$. Finally, we hope that our results can be used as a useful tool to study these thermal properties in connection with experiments. 

\section*{Acknowledgments} 
The authors thank the Conselho Nacional de Desenvolvimento Cient\'{\i}fico e Tecnol\'{o}gico (CNPq), grant n$\textsuperscript{\underline{\scriptsize o}}$ 308638/2015-8 (CASA), and Coordena\c{c}ao de Aperfei\c{c}oamento do Pessoal de N\'{\i}vel Superior (CAPES), for financial support.

\end{document}